\documentclass[aps,prb,twocolumn,showpacs,superscriptaddress,groupedaddress,10pt]{revtex4-2}
\usepackage{graphicx}  
\usepackage{dcolumn}   
\usepackage{bm}        
\usepackage{amssymb}
\usepackage[spanish]{babel}
\usepackage[utf8]{inputenc}
\usepackage{hyperref}
\newcommand{\lr}[1]{\left(#1\right)}

\begin{document}
\title{C\'alculo de la integral escalar a un lazo $A\lr{m^2}$ mediante el teorema del residuo}
\author{J. D. García-Aguilar y J. C. Gómez-Izquierdo}
\affiliation{Centro de Estudios Cient\'ificos y Tecnol\'ogicos No. 16, Instituto Polit\'ecnico Nacional, Pachuca: Ciudad del Conocimiento y la Cultura,
Carretera Pachuca Actopan km 1+1500,\\  San Agust\'in Tlaxiaca, Hidalgo, M\'exico.\\ e-mail: jdgarcia@ipn.mx, cizquierdo@ipn.mx \vspace{10pt}\\
{\small Recibido el 13 de mayo de 2020; Aceptado el 8 de julio de 2020}}

\begin{abstract}
El cálculo de correcciones radiativas en la fenomenología de la física de partículas elementales ha permitido realizar grandes predicciones en las observables físicas del llamado Modelo Estándar, las cuales están en gran acuerdo con las mediciones experimentales de distintos colisionadores y detectores de partículas e incluso a través de predicciones realizadas con métodos numéricos como es el caso de la Cromodinámica Cuántica y su estudio mediante la retícula en el régimen de bajas energías. Este es un trabajo pedagógico en el cual se calcula la integral escalar a un lazo $A\lr{m^2}$, que aparece en los cálculos de correcciones radiativas en el esquema de regularización dimensional. Como resultado principal, se presenta a través de un enfoque no abordado en la literatura una manera de calcular la integral escalar mediante el Teorema del Residuo de variable compleja. Adicionalmente, se demuestra mediante propiedades de la función Gamma la equivalencia de ambos procedimientos.
\\
\\
{\it Descriptores:} Teoría de campo, Renormalización, Propiedades de teoría perturbativa.
 \\
 \\
Radiative corrections in the phenomenology of particle physics lead to great predictions on the observables of the Standard Model (SM) which are in good agreement with different measurements on Particle Accelerators and Detectors and in the case of numeric predictions on Quantum Chromodynamics (QCD) and its study in the lattice for the low energy regime. This is a pedagogical paper in which we calculate the one loop scalar integral $A\lr{m^2}$, that arises in the simplest radiative correction calculation, by means of the dimensional regularization. As main result, we perform also the evaluation of this integral by using the Residue Theorem on complex variable. In addition, the equivalence between the two procedures is shown through Gamma function properties.
\end{abstract}

\maketitle 

\section{Introducción}

Las predicciones teóricas del Modelo Estándar de las Partículas Fundamentales (SM por sus siglas en inglés) no concuerdan a nivel \'arbol con las observables físicas que se pueden obtener a través de mediciones experimentales en colisionadores y detectores de partículas si la precisión experimental es del uno por ciento \cite{Tanabashi:2018oca}, por lo tanto es necesario considerar las correcciones cuánticas llamadas también correcciones radiativas. Estas correcciones radiativas también facilitan obtener información sobre sectores de la teoría que no son directamente observables \cite{Kakuda:2013nva}, de igual manera tienen la capacidad de lograr rupturas de simetrías \cite{Coleman:1973jx} y a la vez son capaces de realizar predicciones en observables físicas como por ejemplo la masa del neutrino \cite{Babu:2002uu}.

Las correcciones radiativas son uno de los objetos de estudio en cualquier curso de teoría cuántica de campos (Quantum Field Theory o QFT por sus siglas en inglés) \cite{Peskin:1995ev} y en el caso más simple, estas correcciones dan origen a las llamadas integrales a un lazo que en el caso del SM toman la forma 
\begin{equation}
I_N\sim\int \frac{d^4 p}{\lr{2\pi}^4}\frac{\mathcal{N}\lr{p}}{\prod_{i=1}^{N} \lr{\lr{p+q_{i-1}}^2-m_i^2}}, \label{intreg}
\end{equation}
donde $N$ es el número de partículas externas cada una con momento $k_i$, $q_j=\sum_{l=1}^j k_l$ y $k_1+k_2+\ldots+k_N=0$ gracias a la conservación del momento\footnote{La relación (\ref{intreg}) es válida para diagramas de Feynman donde cada vértice tiene solo una partícula externa.} \cite{Ellis:2011cr}. En general $\mathcal{N}\lr{p}$ es una función polinomial de $p$ y  el caso de interés en este trabajo es $\mathcal{N}\lr{p}=1$ a las cuales se les conocen como integrales escalares \cite{tHooft:1978jhc}.

Por otra parte, las integrales escalares con una o dos partículas externas divergen en el límite $p\rightarrow\infty$ y por lo tanto es necesario introducir un esquema de \emph{regularizaci\'on}, esto es una, modificación en la teoría que permite aislar los términos divergentes de los  términos que son finitos una vez que se realiza la integración. Existen distintos esquemas de regularización y uno de los más populares es la \emph{regularización dimensional} \cite{Leibbrandt:1975dj} en el que se generaliza la dimensión del espacio-tiempo de cuatro a $D$ dimensiones para resolver la integral y posteriormente hacer uso de una extensión analítica para regresar al espacio cuatro dimensional al utilizar la expresión $D=4-2\varepsilon$, donde $\varepsilon$ es un número complejo llamado regulador. La regularización dimensional finaliza al utilizar el límite $\varepsilon\rightarrow 0$ ya que la divergencia aparece explícitamente en términos de $1/\varepsilon$.


El propósito de este trabajo es presentar, de una forma pedagógica y simple, un método nuevo basado en el Teorema del Residuo dentro del esquema de regularización dimensional para evaluar la integral escalar $A\lr{m^2}$.
Además, se demuestra la equivalencia de dicho método con el cálculo usual en la literatura mediante funciones Beta \cite{Veltman:1994wz}. 

El trabajo está organizado de la siguiente manera. 
En la Sec. \ref{integral} se presenta la integral escalar $A\lr{m^2}$ y su reducción a una integral unidimensional mediante el uso de coordenadas esféricas $D$-dimensionales. En la Sec. \ref{intbeta} se resuelve la integral unidimensional mediante el uso de funciones Beta. En la Sec. \ref{intcontorno} se resuelve la misma integral mediante el Teorema del Residuo, en ambas secciones se muestran los cálculos explícitos. En la Sec. \ref{equiva} se demuestra la equivalencia entre los resultados anteriores. Finalmente, en la Sec. \ref{conclu} se presentan las conclusiones del trabajo. Se ha añadido un apéndice para explicar detalles al lector sobre el resultado principal del trabajo.

\section{La integral $A\lr{m^2}$}\label{integral}

La integral escalar a un lazo $A\lr{m^2}$ es la integral de Feynman más simple que surge en el cálculo de correcciones radiativas  y aparece generalmente al calcular TadPoles \footnote{Un Tadpole o diagrama de renacuajo es un diagrama de Feynman con una sola línea externa, véase \cite{Veltman:1994wz} pag. 149.} o correcciones de masa en modelos de campos escalares \cite{Ramond} la cual se define en la forma
\begin{equation}
A\lr{m^2}=\int d^D p\frac{1}{\mathbf{p}^2+m^2}, \label{int}
\end{equation}
donde $\mathbf{p}$ es un $D$-vector con norma Euclidiana
\begin{equation}
\mathbf{p}^2=\sum_{k=1}^D p_k^2,
\end{equation}
y $m$ la masa de un campo asociado al proceso.

Esta integral escalar  $D$-dimensional  puede reducirse a una integral unidimensional mediante cambios de variable de coordenadas cartesianas a coordenadas esféricas en $D-$dimensiones 
\begin{equation}
\lr{p_1,\ldots, p_D}\rightarrow \lr{p,\phi_1,\ldots,\phi_{D-2},\theta},
\end{equation}
 con las reglas de transformación \cite{Blumenson:1967} 
\begin{eqnarray}
p_1&=&p\cos\lr{\phi_1},\nonumber\\
p_j&=&p\cos\lr{\phi_j}\prod_{k=1}^{j-1}\sin\lr{\phi_k}\hspace{0.5cm}\lr{j=2,\ldots,D-2},\nonumber\\
p_{D-1}&=&p\sin\lr{\theta}\prod_{k=1}^{D-2}\sin\lr{\phi_k},\nonumber\\ 
p_D&=&p\cos\lr{\theta}\prod_{k=1}^{D-2}\sin\lr{\phi_k}, \label{rule}
\end{eqnarray}
donde $\phi_j\in \left[0,\pi\right]$ para $j=1,\ldots,D-2$; $\theta\in\left[0,2\pi\right]$ y $p\in\left[0,\infty\right]$. Estas reglas de transformación (\ref{rule}) permiten escribir al diferencial de la integral (\ref{int}) en la forma
\begin{equation}
d^Dp=p^{D-1} dp d\theta\prod_{k=1}^{D-2}\sin\lr{\phi_k}d\phi_k.
\end{equation}

De acuerdo a las definiciones anteriores es posible reescribir la integral (\ref{int}) en la forma
\begin{eqnarray}
A\lr{m^2}=S_D\int_0^\infty \frac{p^{D-1} dp}{p^2+m^2}, \label{expres}
\end{eqnarray}
donde $S_D$ es la superficie (área) de una esfera $D$-dimensional de radio uno que surge de la integral 
\begin{equation}
S_D=\int d\theta\prod_{k=1}^{D-2}\sin\lr{\phi_k}d\phi_k,
\end{equation}
la cual puede calcularse mediante fórmulas recursivas a partir del volumen de dicha esfera $D$-dimensional \cite{Sommerville} y con ello llegar a la expresión 
\begin{equation}
S_D=\frac{2\pi^{D/2}}{\Gamma\lr{D/2}},
\end{equation}
donde $\Gamma\lr{z}$ es la función Gamma que cumple la condición $z\Gamma\lr{z}=\Gamma\lr{z+1}$ \cite{Arfken:2013}.

\section{La integral $A\lr{m^2}$ mediante funciones Beta}\label{intbeta}

En la sección anterior se demostró que la solución a la integral escalar (\ref{int}) se reduce a resolver la integral
\begin{equation}
I\lr{m^2}=\int_0^\infty \frac{p^{D-1}}{p^2+m^2}dp, \label{intdos}
\end{equation}
la cual requiere utilizar una continuación analítica \cite{Arfken:2013} debido a que solo es convergente para valores de $D<2$.

El cambio de variable
\begin{equation}
p^2=y m^2\Rightarrow dp=\frac{m^2 dy}{2p},
\end{equation}
permite escribir la integral (\ref{intdos})  en la forma
\begin{eqnarray}
I\lr{m^2}&=&\int_0^\infty \frac{p^{D-1}}{p^2+m^2}dp\nonumber\\
&=&\frac{m^{D-2}}{2}\int_0^\infty \frac{y^{D/2-1}}{1+y}dy\nonumber\\
&=& \frac{m^{D-2}}{2} B\lr{\frac{D}{2},1-\frac{D}{2}}\nonumber\\
&=& \frac{m^{D-2}}{2}\frac{\Gamma\lr{\frac{D}{2}}\Gamma\lr{1-\frac{D}{2}}}{\Gamma\lr{1}},
\end{eqnarray}
donde se utilizó la definición de la función Beta
\begin{equation}
B\lr{p+1,q+1}=\int_0^\infty \frac{u^p}{\lr{1+u}^{p+q+2}}du,
\end{equation}
 y la solución de dicha integral en términos de las funciones Gamma \cite{Arfken:2013} está dada por
 \begin{equation}
 B\lr{p,q}=\frac{\Gamma\lr{p}\Gamma\lr{q}}{\Gamma\lr{p+q}}.
 \end{equation}
 
 Por lo tanto, la integral escalar está dada por la expresión
 \begin{equation}
 A\lr{m^2}=\frac{\pi^{D/2}m^{D-2}}{\Gamma\lr{D/2}}\lr{{\Gamma\lr{\frac{D}{2}}\Gamma\lr{1-\frac{D}{2}}}}, \label{betafunction}
 \end{equation}
 la cual depende explícitamente del parámetro $m$ y de la dimensión $D$ que se está considerando.

\section{La integral $A\lr{m^2}$ mediante el teorema de residuo}\label{intcontorno}
La integración por contornos en el plano complejo es uno de los métodos más convenientes para la evaluación de integrales definidas y se sustenta como una consecuencia del Teorema del Residuo en análisis complejo \cite{Mathews}. Este teorema menciona que si una función $f\lr{z}$ es regular en una región acotada por un contorno cerrado $C$ excepto para un número finito de polos en el interior de dicho contorno, entonces la integral de $f\lr{z}$ a lo largo del contorno $C$ es
\begin{equation}
\oint_C f\lr{z} dz=2\pi i\sum Res\lr{f\lr{z},z_0}, \label{Teorema}
\end{equation} 
donde $Res\lr{f\lr{z},z_0}$ representa el residuo de la función en la singularidad aislada $z=z_0$ (en \cite{Schaum} se muestran ejemplos de como obtener dichos residuos). Por lo tanto el problema de evaluar integrales de contorno se reemplaza por el problema algebraico de calcular los residuos en los puntos singulares de una función. 

Es de notar que la integral (\ref{intdos}) representa la integral de la  función
\begin{equation}
f\lr{z}=\frac{z^{D-1}}{z^2+m^2}, \label{function}
\end{equation} 
sobre el eje real positivo en el plano complejo. Esta función tiene polos en los puntos $z=\pm im$ y es analítica en el resto del plano y por lo cual su solución con el uso del Teorema del Residuo implica considerar la integral
\begin{equation}
\oint_C f\lr{z} dz, \label{fuente}
\end{equation}
\begin{figure}[t]
\includegraphics[scale=0.65]{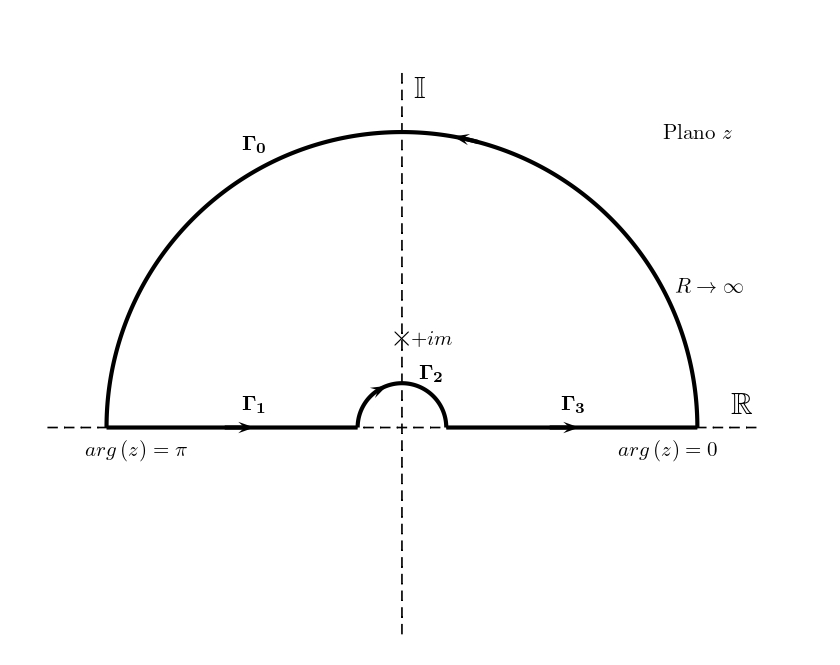}
\caption{Contorno para evaluar la integral que consta del semicírculo $\Gamma_0$ de radio $R$, el eje real $\Gamma_1$ desde $-R$ hasta $-\epsilon$, el semicírculo $\Gamma_2$ de radio $\epsilon$ y el eje real $\Gamma_3$ desde $\epsilon$ hasta $R$.}
\label{contorno}
\end{figure}
a lo largo del contorno de la Fig. \ref{contorno} y reescribirla como (v\'ease Ap\'endice \ref{cambiosvar})
\begin{equation}
\oint_C f\lr{z}dz=\lr{1-\lr{-1}^D}\int_0^\infty\frac{t^{D-1}}{t^2+m^2}dt. \label{intfinal}
\end{equation}

Por otra parte, el residuo de la función (\ref{function}) en el polo $z=im$ \cite{Schaum} (ya que es el único dentro del contorno) es
\begin{eqnarray}
z_1&=&\lim_{z\rightarrow im}\lr{z-im}f\lr{z} \nonumber\\
&=&\lim_{z\rightarrow im}\frac{z^{D-1}}{z+im}=\frac{\lr{im}^{D-1}}{2im}=-\frac{\lr{im}^D}{2m^2}.\label{residuos}
\end{eqnarray}

Finalmente, de acuerdo a la Ec. (\ref{Teorema}) y los resultados de (\ref{intfinal}) y (\ref{residuos}) se encuentra que
\begin{equation}
I\lr{m^2}=-\frac{2\pi i}{\lr{1-\lr{-1}^D}}\lr{\frac{(i m)^D}{2 m^2}},
\end{equation}
y en consecuencia la integral escalar toma la forma
\begin{equation}
 A\lr{m^2}=\frac{\pi^{D/2}m^{D-2}}{\Gamma\lr{D/2}}\lr{2\pi i\frac{-i^D}{1-\lr{-1}^D}}, \label{contornofunction}
 \end{equation}
 que de igual manera a la Ec. (\ref{betafunction}) tiene la dependencia explícita en $D$ y $m$.
 
\section{Equivalencia de resultados}\label{equiva}
En esta sección se  demuestra la equivalencia de las Ecs. (\ref{betafunction}) y (\ref{contornofunction}) mediante el uso del álgebra de números complejos, la fórmula de reflexión de la función Gamma \cite{Arfken:2013}
\begin{equation}
\Gamma\lr{1-z}\Gamma\lr{z}=\frac{\pi}{\sin\lr{\pi z}},
\end{equation}
y las igualdades
\begin{eqnarray}
e^{i\theta}&=&\cos\lr{\theta}+i\sin\lr{\theta},\nonumber\\
\sin\lr{2\theta}&=&2\cos\lr{\theta}\sin\lr{\theta},\nonumber\\
\sin^2\lr{\theta}&=&\frac{1}{2}\lr{1-\cos\lr{2\theta}}.
\end{eqnarray}

A partir de lo anterior
\begin{eqnarray}
2\pi i\frac{-i^D}{1-\lr{-1}^D}&=&2\pi i\frac{-e^{i \pi/2 D}}{1-e^{i\pi D}}\\
&=&2\pi i\frac{-\lr{\cos\lr{\frac{\pi D}{2}}+i\sin\lr{\frac{\pi D}{2}}}}{1-\cos\lr{\pi D}-i\sin\lr{\pi D}}\nonumber\\
&=&2\pi i\frac{-\lr{\frac{\sin{\pi D}}{2\sin\lr{\frac{\pi D}{2}}}+i\frac{\sin^2\lr{\frac{\pi D}{2}}}{\sin\lr{\frac{\pi D}{2}}}}}{1-\cos\lr{\pi D}-i\sin\lr{\pi D}}\nonumber\\
&=&\frac{\pi}{\sin\lr{\frac{\pi D}{2}}}\frac{-i\sin\lr{\pi D}+1-\cos\lr{\pi D}}{1-\cos\lr{\pi D}-i\sin\lr{\pi D}}\nonumber\\
&=&\frac{\pi}{\sin\lr{\frac{\pi D}{2}}}=\Gamma\lr{1-\frac{D}{2}}\Gamma\lr{\frac{D}{2}}.\nonumber
\end{eqnarray}
con lo que se demuestra que ambos caminos de integración conducen al mismo resultado.

\section{Conclusiones}\label{conclu}

En este trabajo se realizó la evaluación de la integral escalar $A\lr{m^2}$ por dos métodos distintos. El primero de ellos con el método usual en la literatura mediante funciones Beta, mientras que el segundo hace uso del Teorema del Residuo de variable compleja para una integración por contornos. Adicionalmente, utilizando algunas propiedades de la función Gamma existentes en libros de texto de métodos matemáticos, se muestra que ambos resultados son equivalentes. 

Los cálculos aquí presentados son útiles para su discusión en cursos de funciones especiales e introductorios de teoría cuántica de campos y pueden extenderse para otras integrales escalares que emanen de diagramas de Feynman con dos ó más propagadores, un ejemplo son los diagramas triangulares comunes en el estudio de anomalías.

\appendix

\section{Cambios de variable}\label{cambiosvar}

La solución a la integral (\ref{fuente}) a lo largo del contorno de la Figura \ref{contorno} puede  llevarse a cabo considerando dos semicírculos ($\Gamma_0$ y $\Gamma_2$) y dos semirrectas ($\Gamma_1$ y $\Gamma_3$) con lo cual
\begin{eqnarray}
\oint_C f\lr{z}dz&=&\int_{\Gamma_0} f\lr{z}dz+\int_{\Gamma_1} f\lr{z}dz\nonumber\\&&+\int_{\Gamma_2} f\lr{z}dz+\int_{\Gamma_3} f\lr{z}dz,
\end{eqnarray}
y por lo tanto para cada camino de integración es posible utilizar distintos cambios de variable.

Para el caso de la trayectoria $\Gamma_0$, se propone utilizar el cambio de variable $z=Re^{i\theta}$ con $\theta\in\left[0,\pi\right]$ y con ello
\begin{eqnarray*}
dz&=&i R e^{i\theta}d\theta; \\
\frac{z^{D-1}}{z^2+m^2}&\approx& R^{D-3} e^{i\theta\lr{D-3}};\\
\int_{\Gamma_0} \frac{z^{D-1}}{z^2+m^2} dz&\approx&i\int_0^\pi \frac{e^{i\theta\lr{D-2}}}{R^{2-D}}d\theta \rightarrow 0 \mbox{ si } R\rightarrow\infty,
\end{eqnarray*}
donde aquí vale la pena recalcar que el resultado anterior es válido si y solo si $D<2$ y por lo tanto para cualquier valor de $D>2$ es necesario hacer uso de una extensión analítica \cite{Arfken:2013} para obtener un resultado convergente.

En la trayectoria $\Gamma_1$ es posible escribir $z=te^{i\pi}$ donde $t\in\lr{R,\epsilon}$, con lo cual
\begin{eqnarray*}
dz&=&e^{i\pi}dt;\\
\int_{\Gamma_1} \frac{z^{D-1}}{z^2+m^2}dz&=&e^{i\pi D }\int_{R}^{\epsilon} \frac{t^{D-1}}{t^2+m^2}dt\\
&=&-\lr{-1}^D\int_\epsilon^{R} \frac{t^{D-1}}{t^2+m^2}dt.
\end{eqnarray*}

El cálculo sobre la trayectoria $\Gamma_2$ es similar al realizado sobre $\Gamma_0$ utilizando el cambio de variable $z=\epsilon e^{i\theta}$ con $\theta \in \left[\pi,0\right]$
\begin{eqnarray*}
dz&=&i \epsilon e^{i\theta}d\theta; \\
\frac{z^{D-1}}{z^2+m^2}&\approx& \epsilon^{D-1} \frac{e^{i\theta\lr{D-1}}}{m^2};\\
\int_{\Gamma_2} \frac{z^{D-1}}{z^2+m^2} dz&\approx&i\epsilon^D\int_\pi^0 \frac{e^{i\theta D}}{m^2}d\theta \rightarrow 0 \mbox{ si } \epsilon\rightarrow 0,
\end{eqnarray*}
donde, a diferencia del caso para $\Gamma_0$, este resultado no requiere una extensión analítica.

Finalmente, sobre la trayectoria $\Gamma_3$ el cálculo es directo si se utiliza $z=t$  en el intervalo $t\in\lr{\epsilon, R}$  
\[
\int_{\Gamma_3} \frac{z^{D}}{z^2+m^2}dz=\int_\epsilon^R \frac{t^{D-1}}{t^2+m^2}dt.
\]

Los resultados anteriores muestran que en los límites $R\rightarrow\infty$ y $\epsilon\rightarrow 0$ la integral (\ref{fuente}) se reduce a la expresión
\begin{equation}
\oint_C f\lr{z}dz=\lr{1-\lr{-1}^D}\int_0^\infty \frac{t^{D-1}}{t^2+m^2}dt.
\end{equation}

\subsection*{Agradecimientos}
García-Aguilar agradece las facilidades dadas por el IPN a través del proyecto SIP número 20201313. J. C. Gómez-Izquierdo agradece al Instituto Politécnico Nacional
(IPN) por ser beneficiado con el proyecto SIP 20202024.

\end{document}